\documentclass[12pt,preprint]{aastex}

\begin{document}

\title{A {\it Chandra} Observation of a TW Hydrae Association Brown Dwarf}

\author{John E. Gizis}
\author{Ravi Bharat}
\affil{Department of Physics and Astronomy, University of Delaware,
Newark, DE 19716}

\begin{abstract}     
We present {\it Chandra} observations of the young brown dwarf
2MASSW J1207334-393254, which is a probable member of the TW Hya association.
Although this substellar object has strong H$\alpha$ emission, it
has no detected X Ray flux in a fifty kilosecond ACIS-S observation.  We place
a conservative 
upper limit of $1.2 \times 10^{26}$ erg/sec on its X-ray luminosity.
We compare our M8 target to the similar mass object 
TWA 5B, which has weaker H$\alpha$ emission but strong X-ray emission.  
We argue our results are consistent with the notion that 
2MASSW J1207334-393254 is interacting with a disk.
\end{abstract}

\keywords{stars: low-mass, brown dwarfs --- open clusters and associations: individual (TW Hya Association) --- stars: activity}

\section{Introduction}

Brown dwarfs have proven to show a variety of phenomena that 
can be attributed to magnetic activity.  Amongst the field 
late-M and L dwarfs, H$\alpha$ activity declines sharply
as a function of effective temperature for both stars and
brown dwarfs \citep{g00}. In contrast to those relatively old 
($>100$ Myr) field dwarfs, young ($<10$ Myr) brown dwarfs
show a wide range of behavior.  Some of these youngest
brown dwarfs show strong H$\alpha$ emission; others
show strong X-ray emission \citep[among others]{safe,ic348,rhooph,feig}.  
Most notably, high H$\alpha$
emission has been viewed a possible sign of accretion,
as in classical T Tauri stars \citep{muze,bnm03,xtwa5b,
mohanty}.  
Since the nearest star-forming regions are more than
a hundred parsecs away, detailed observations of the
faintest brown dwarfs is difficult.  

At a distance of only fifty parsecs, the TW Hydrae Association offers 
the opportunity to study a few brighter young brown dwarfs.  The
age is estimated to be $\sim 10$ Myr \citep{webb99,mf01}.  While most of the
few dozen members are pre-main sequence stars, three brown dwarf
members have been detected.  TWA 5B, an M9 dwarf, was detected
as a companion to more massive primary\citep{twa5b,hatwa5b}. 
2MASSW J1207334-393254 (hereafter 2M1207) and the candidate
2MASSW J1139511-315921 were identified by \citet{g02} using the 
Two Micron All-Sky Survey (2MASS).
\footnote{These sources appear as 
2MASS J12073346-3932539 and 2MASS J11395113-3159214 in the final 2MASS 
release.}  Both 2M1207 and 2M1139 are classified as spectral type M8. 
TWA5B was detected as an X-ray source by \citet{xtwa5b}.
In this letter, we report {\it Chandra} observations of 2M1207.    

\section{Data Analysis}

\subsection{Optical Data Revisited}

The original G02 spectrum showed extremely strong H$\alpha$ emission
of $\sim 300$\AA~ in two consecutive spectra.  (We report
emission line strengths as pseudo-equivalent widths throughout this
letter). 
More recent high resolution spectra by \citet{mohanty} showed emission of
28 \AA -- still strong, but considerably weaker.  Kirkpatrick (2003,
priv. comm.) has kindly provided a Keck LRIS spectrum which shows
42\AA~ emission.  We conclude that the G02 observation captured
either a flare or some other transient activity state; 2M1207
is certainly variable in its H$\alpha$ strength but all observations
show strong emission.  The Kirkpatrick spectrum confirms the
low surface gravity and hence young age of 2M1207; \citet{mohanty}
conclude that the radial velocity is consistent with membership.
\citet{mohanty} find that the H$\alpha$ is broad and conclude
that 2M1207 is (weakly) accreting from a disk.  Other than the
emission lines, the two spectra appear very similar and we conclude
there is no sign of veiling during the high activity state.  

G02 used marginal detections on photographic plates with the
2MASS data to argue that the observed proper motion is
small and consistent with TWA membership.  In an attempt to
confirm this result, Tinney (priv. comm.) obtained a J-band image
for us using the AAO IRIS2 camera on 11 June 2003.  
We measured the position of 
2M1207 with respect to other stars in the field and compared to
the 10 May 1997 2MASS data.  These measurements find that 
2M1207 is in motion to the west (63 mas/yr) and south (40 mas/yr) 
as expected, but judging by the residuals in individual field stars
we estimate our measurement uncertainties to be in the range
50-100 mas/yr.  Thus, these results again support a small
motion and a young age, but are not adequate to definitively 
confirm association membership.  The DENIS survey 
detected 2M1207 with $I=15.88 \pm 0.06$ \citep{denis}.
The IJHK colors are normal for an M8 dwarf and do not appear
to be reddened.

\subsection{X Ray Data}

Chandra obtained a 51 ksec observation of 2M1207 in March 2003
using ACIS-S.  Examination of the image reveals that no source is
found at the position of 2M1207, with the maximum pixel
within 5 arcseconds having a value of only two (2) counts.  
Pixels with two counts are common throughout the image, and overall
the characteristics of the pixels near 2M1207's position is
the same as that in other comparison (sky) regions throughout
the image.  We conclude that we have not detected 
2M1207 in X rays, either in a quiescent or flaring state.  
We set a conservative upper limit of six photons, three times the
maximum observed sky excursions, as our upper limit on
the observed counts.  

Since TWA 5B has been detected in X-rays with the same instrument,
we use it to set the upper limits on 2M1207's X rays.
\citet{xtwa5b} had 35 photons detected in a 0.7 arcsecond
aperture in 9.3 ksec.  Assuming the same distance for 2M1207,
we find
$$L_X = \frac{N}{35}\frac{9.3 \rm{ksec}}{51 \rm{ksec}}
4\times10^{27} \rm{erg/sec}$$
Taking $N=6$ photons, we find $L_X < 1.2 \times 10^{26}$ erg/sec.
2M1207 is actually slightly fainter than TWA 5B ($\Delta H = 0.3$),
which might be due to a lower bolometric luminosity.
In this case, case we scale from TWA 5B's 
$\log L_X/L_{Bol} = -3.4$ to 2M1207: $\log L_X/L_{Bol} < -4.8$.

\section{Analysis}

We wish to determine whether the faint X-ray luminosity
is unexpected in light of the strong H$\alpha$ emission.   
First, we compare 2M1207 to field M8/M9 dwarfs.  The data for field
dwarfs are sparse, but suggest that both the X-ray coronal flux
and the C{IV} ultraviolet transition-region flux is proportional
to the H$\alpha$ chromospheric flux \citep{vb10,stis}, as
in earlier-type M dwarfs. In other words, we assume the strength of the
chromosphere is proportional to the strength of the corona and
the two are physically related.    
We can therefore look to see whether 2M1207 has 
a temperature inversion resulting in a chromosphere
and corona.  
Deep X-ray data are available for a few field M8/M9
dwarfs: LHS 2065 \citep{lhs2065}, VB10 \citep{vb10},
LHS 2924 \citep{lhs2924}, BRI 0021 \citep{neu99}
and LP 944-20 \citep{rutledge,mb2002}.  X-ray activity like
LHS 2065 ($\log L_X/L_{Bol}\approx-3.4$, H$\alpha =8-25$\AA~,
Mart{\'{\i}}n \& Ardila 2001)
can be easily ruled out, and 
VB10-like ($\log L_X/L_{Bol}= -4.9$)
activity levels lie near our upper limits; its H$\alpha$ emission
is only 4.4\AA~ \citep{tinneyreid}.
The existing upper limit on the M9 dwarf LHS 2924
($\log L_X/L_{Bol}< -4.3$, H$\alpha = 4$\AA) is consistent with 
the deeper VB10 detection 
LP 944-20's activity level ($\log L_X/L_{Bol}< -6.2$)
is well below our 2M1207 limit, but it only has
$1.2$\AA~ H$\alpha$ emission \citep{tinney}.  
BRI0021 is similar to  
LP 944-20 in that it is also rapid rotator with very
weak or no H$\alpha$ emission \citep{bm95,tinneyreid}; its limit of
$\log L_X/L_{Bol} < -4.68$  is consistent with the deeper LP 944-20
data.   
Plainly, if young brown dwarfs behave like similar temperature, 
older field brown dwarfs, then the observed 2M1207 H$\alpha$ emission
should result in X-ray activity well above our limits.  
Turning the model around, the observed X-ray limits suggests
that less than $\sim 5$\AA~ of H$\alpha$, or $<20$\% of
the observed emission, can be due to a traditional chromosphere.  
The bulk of the observed H$\alpha$ emission is then due 
to some non-chromospheric process.  

Brown dwarfs in a number of young $<10$ Myr star-forming regions
have been detected by {\it Chandra} \citep{ic348,rhooph,feig}.  The detections
have generally been at the $\log L_X/L_{Bol} \approx -3$ level, while 
many other brown dwarfs remain undetectable in X rays.  
\citet{xtwa5b} point out that the X-ray luminous brown dwarfs 
all have H$\alpha \lesssim 25-30$\AA; those with H$\alpha \gtrsim 25$\AA~
are all undetected in X rays.  This is consistent with the
fact the TWA5B was detected but 2M1207 was not.  
\citet{xtwa5b} also point out that that in the X-ray detected
objects, the H$\alpha$ and the X-rays are correlated, and
argue this supports a chromospheric/coronal relation.  
Their correlation applied to the 2M1207 upper limit 
suggests that $<2.2$\AA~ of 2M1207's H$\alpha$ emission can be attributed to
chromosphere.  This is remarkably consistent with our results
from the field dwarf comparison.  By analogy with 
classical T Tauri stars, they attribute the high H$\alpha$ 
activity to star-disk interactions.  Our X-ray data supports
\citet{mohanty}'s conclusion that the broad H$\alpha$ profile
of 2M1207 is due to accretion.  One counterargument is that 
2M1207 does not show an infrared excess \citep{jay}, but
we note that \citet{muze} argued that H$\alpha$ accretion 
remains detectable in substellar objects while UV/optical/IR excesses
do not.

We regard our results as consistent with the emerging picture
that some young brown dwarfs show high activity due to star-disk
interactions.  Since 2M1207 is $\sim 10$ Myr old, evidently
some brown dwarfs have fairly long-lived disks 
and ongoing accretion, or at least interactions.  It is not clear whether the 
presence of the disk somehow suppresses the chromosphere/corona
of young brown dwarfs; indeed we do not have an observable
parameter that correlates with the presence or absence of 
a corona.  Deeper X ray data in star-forming regions may be
needed to characterize the brown dwarfs with disks.  

Finally, we note that TWA5B is probably variable in H$\alpha$,
since \citet{mohanty} report 5.1\AA~ but \citet{hatwa5b}
report 20\AA~ of emission.  In this regard, it may well
be similar to LHS 2065, which shows similar levels of variability 
\citep{ma2001} and has a similar X-ray activity level.  
Since both TWA5B and 2M1207 have been observed to be quite variable,
the more distant brown dwarfs in PMS clusters need to also be
monitored in H$\alpha$ and X rays (if feasible) 
before robust conclusions on activity can be drawn.
Still, the similarity of LHS 2065 and TWA5B is suggestive that
chromospheres and coronae of objects with the same spectral types 
(photospheric temperatures) may be quite similar 
despite their different ages.  

\section{Conclusion}

We report an upper limit to the X-ray activity level of the
young brown dwarf 2M1207. Comparison to both field and
young brown dwarfs suggests that this source must then
have a weak chromosphere.  We attribute the strong observed
H$\alpha$ emission to star-disk interactions.  Activity in 
other objects such as TWA5B seems to be similar to the
most active field M8-M9 dwarfs, and can be attributed to
chromospheres and coronae.  
2M1207 is the closest known accreting brown dwarf and
deserves further study at all wavelengths.  
We are scheduled to observe 2M1207 in the ultraviolet using
Hubble Space Telescope to search for a transition region, which
should be weak if our interpretation of the {\it Chandra} data
is correct.  

We thank Marc Audard for conversations about Chandra data and
activity in brown dwarfs, Chris Tinney for obtaining 
AAO data, and Davy Kirkpatrick for sharing his unpublished Keck
spectrum.     
We also thank the {\it Chandra} team
for the development of this powerful satellite.  
Support for this work was provided by the National Aeronautics and
Space Administration through Chandra award number GO-4020X issued
by the Chandra X-ray Observatory Center, which is operated by 
the Smithsonian Astrophysical Observatory for and on behalf of 
the National Aeronautics and
Space Administration under contract number NAS8-39073


\end{document}